**Josh Simon** is an astronomer at the Carnegie Observatories in Pasadena, California. **Marla Geha** is a professor in the astronomy and physics departments at Yale University in New Haven, Connecticut.

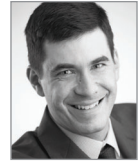
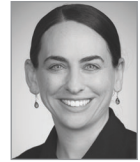

# Illuminating the
# DARKEST GALAXIES

Joshua D. Simon and Marla Geha

**The behavior of extremely dim galaxies provides stringent constraints on the nature of dark matter. Establishing those constraints depends on precise stellar-motion measurements.**

You might not think that galaxies and Hollywood celebrities have much in common. But like a true celebrity, our Milky Way is surrounded by a galactic entourage. We currently know of roughly 60 smaller galaxies in orbit around it, and an equal or greater number are thought to remain undiscovered. And like a good entourage, the satellites have a combined brightness that is less than the Milky Way itself by an order of magnitude. The smallest and most numerous of the satellites are known as ultrafaint dwarf galaxies, and both their number and their internal structures provide crucial information about the nature of dark matter.

Although we cannot directly observe dark matter, stellar motions in a galaxy are dictated by its gravitational potential and can reveal the dark matter's spatial distribution. Evidence of dark matter is observed in nearly all galaxies, but ultrafaint galaxies are the most extreme, with dark matter typically making up approximately 99.9% of their mass.[1] The meager 0.1% of ordinary matter is also exceptional—the majority of stars in ultrafaint dwarf galaxies were formed shortly after the Big Bang, making them unique probes of stellar nucleosynthesis (see the article by Anna Frebel and Timothy C. Beers, PHYSICS TODAY, January 2018, page 30) and galaxy formation. Yet those valuable galaxies were unknown before 2005. It turns out they were hiding in plain sight.

Through most of the 20th century, searches for new dwarf galaxies around the Milky Way relied on visual examination of photographic plates. Surveys such as the Palomar Observatory Sky Survey and the European Southern

Observatory and Science Research Council's Southern Sky Survey imaged the entire sky using small telescopes equipped with the latest in photographic technology. At best, the hypersensitized glass plates used at observatories could record photons with about 3% efficiency. Nevertheless, a handful of dwarf galaxies were identified through careful inspection (by eye, with a magnifying glass!) of the survey images. By 2000 the known population of dwarf galaxies orbiting the Milky Way totaled 11 (see figure 1), none of which were ultrafaint. The brightest two, the Magellanic Clouds, have luminosities approximately a billion times that of the Sun and are visible to the naked eye in the southern sky. The faintest, in contrast, is the Sextans dwarf spheroidal. Discovered in 1990, it has a luminosity equivalent to that of just 300 000 suns.[2]

The invention of the CCD in 1969 enabled the era of modern astronomy. CCD detectors record nearly all incident photons and in minutes can reveal faint stars that remained invisible in hour-long exposures with photographic plates. Faster and more efficient detectors were developed throughout the 1980s and 1990s, and similar digital detectors are now in nearly all our pockets in the form of camera phones (see PHYSICS TODAY, December 2009, page 12). But it took time for semiconductor manufacturers to build sufficiently large devices to compete with photographic plates for imaging large areas — a square degree or more — of the sky.

Modern digital surveys began in the early 2000s with the Sloan Digital Sky Survey and continue with, among others, the Dark Energy Survey (see the article by Joshua Frieman, PHYSICS TODAY, April 2014, page 28). Digital imaging now covers almost the whole sky and records objects an order of magnitude dimmer than those found in the best photographic surveys. The images are captured electronically, so catalogs of the position, brightness, and color of every detected star or galaxy are generated automatically. Rather than poring over physical images to identify the fuzzy, tiny patch of a faint galaxy, astronomers can use computer algorithms to isolate groups of stars with the appropriate brightness and color to reside in a dwarf galaxy.

The digital surveys almost immediately revealed dwarf-galaxy candidates — groups of stars that appeared to be located at tens to hundreds of kiloparsecs from the Sun. The faintest of the newly discovered objects consist of merely 1000 stars, thus earning their "ultrafaint" moniker.

Images alone, however, cannot confirm that those objects are indeed galaxies. The stars must be gravitationally bound to each other rather than be a chance alignment of unrelated stars at different distances (see box 1). To establish the system's nature, its dynamical mass — the mass inferred from the motions of its stars — must be measured and compared with the total mass of the stars. If those values are equal, then the collection is considered a star cluster. But if the dynamical mass is much larger, it's a galaxy. Demonstrating that a candidate meets that criterion requires measuring the velocities of a substantial number of stars in each group.

## Are they really galaxies?

On a warm evening in February 2007, the two of us were on the island of Hawaii. We were preparing to use the 10-meter Keck II Telescope — along with Keck I, the two most powerful optical telescopes in the world — to measure the motions of stars in the first candidate dwarf galaxies discovered by CCD observations. Eight such objects had recently been published by Sloan Digital Sky Survey researchers.[3,4] If any candidates were indeed galaxies, they would be the first new Milky Way satellites discovered in more than a decade. Confirming all eight candidates would nearly double the known population of Milky Way satellites. Between us, we optimistically hoped that one or two of the candidates would turn out to be real. But just a few days before our scheduled telescope time, the weather forecast suggested that cloudy skies would ruin all three nights of our observing run.

We planned to use the telescope's Deep Extragalactic Imaging Multi-Object Spectrograph (DEIMOS) to obtain spectra of stars in the candidate galaxies. Stellar spectra contain dark absorption lines at fixed wavelengths, signatures of thermally generated photons from the hot stellar interior that are being absorbed by the star's cooler photosphere. By measuring the shift in those lines from their expected rest wavelengths—their Doppler shifts—we can determine the line-of-sight, or radial, velocity of the star.

**FIGURE 1. SATELLITE GALAXIES** are easier to discover since the advent of digital sky surveys. **(a)** Before 2000, only 11 were found, largely by scouring photographic plates. The Sloan Digital Sky Survey (SDSS) and the Dark Energy Survey and others (DES+) are responsible for increasing that number to more than 60. **(b)** The spatial distribution of satellite galaxies around the Milky Way.

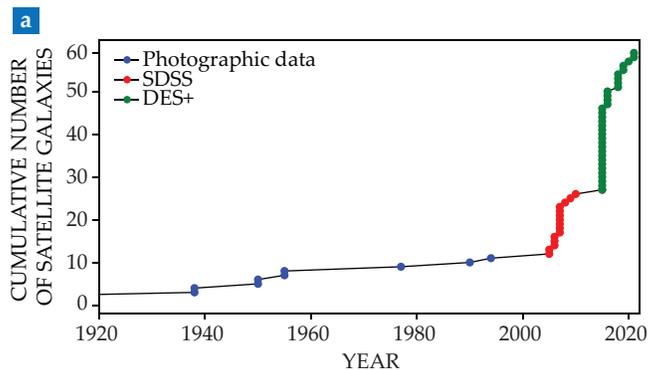
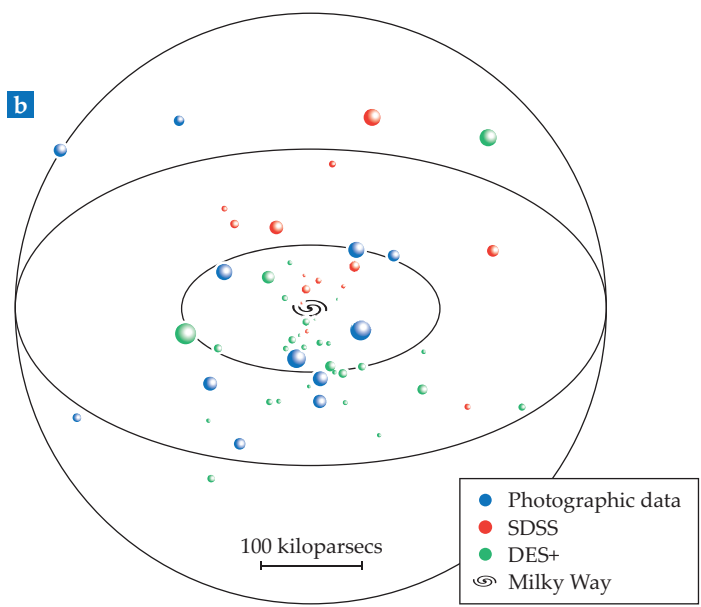

The accuracy of a radial-velocity measurement is set by the accuracy to which the centers of the absorption lines can be determined. Higher-resolution spectrographs and longer exposure times can both generate more accurate radial velocities. DEIMOS can obtain simultaneous spectra for nearly 200 stars with an accuracy of 2 km/s. Stars in any given dwarf galaxy share approximately the same radial velocity, with a variation of 5–10 km/s (see figure 2). Stars that do not belong to the dwarf galaxy, typically foreground stars residing in the Milky Way, can span a wide range of velocities (about 100 km/s). Thus DEIMOS has sufficient resolution to determine whether a star is gravitationally bound to one of the Milky Way's dwarf galaxies or is instead part of the Milky Way itself.

The same stellar-velocity data can also be used to determine the mass of a dwarf galaxy. Once a sample of stars is identified as belonging to a dwarf galaxy, the galaxy's total mass can be computed from the velocity dispersion (the width of the velocity distribution) using Newton's law of gravitation. Although the first dwarf-galaxy velocity-dispersion estimates, obtained in the 1980s by Marc Aaronson[5] and others, relied on single-digit numbers of stars, the minimum number of member stars needed to confirm a dwarf-galaxy candidate is usually larger than 10. Improvements in telescopes, spectrographs, and detector technology have now made it possible to measure velocities for up to a few thousand stars in the largest dwarfs. The resulting mass and density determinations are crucial for dark-matter experiments.

On the eve of our 2007 Keck observing run, the weather made a welcome turnaround. During our three nights, we measured radial velocities for more than 1000 stars and confirmed that all eight of the candidates we targeted were in fact dwarf galaxies. Since then, more than 35 similar systems have also been confirmed by stellar spectroscopy.[6]

Remarkably, our observations also suggested that the total masses of those faint galaxies were overwhelmingly dominated by dark matter rather than their visible stars. The known Milky Way dwarf galaxies already shared that property, but the ultrafaint dwarf galaxies are more extreme, with visible stars accounting for well under 1% of the total mass. The ultrafaint systems therefore provide excellent laboratories for testing theories of dark matter. Below we describe three tests: the overall number of dwarf galaxies around the Milky Way, the amount of dark-matter annihilation radiation from each galaxy, and the galaxies' internal density structures. Each test provides unique information about dark matter.

## Counting dark-matter halos

The prevailing cosmological model, developed over the past 40 years, is based on the concept of cold dark matter (CDM). In that context, "cold" refers to the typical velocity of dark-matter particles—they would have been nonrelativistic when they decoupled from baryons in the early universe. More recently, the discovery of dark energy has motivated the ΛCDM model, which includes both CDM and a dominant cosmological-constant term (Λ). The cosmological constant accounts for the acceleration in the expansion rate of the universe (see PHYSICS TODAY, December 2011, page 14), but it does not affect dynamics at the scale of individual galaxies.

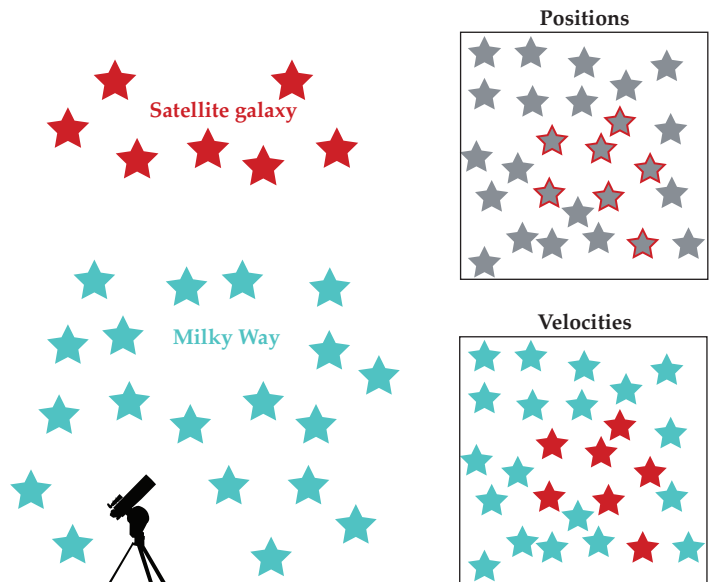

**FIGURE 2. A SATELLITE GALAXY** imaged by an Earth- or space-based telescope, as illustrated on the left, captures only the stars' positions. That information (top right) is insufficient to definitively determine which stars in the field of view are part of the satellite galaxy. But the satellite is moving relative to the Milky Way, where the observer is located. When velocity data are taken into account (bottom right), the stars become distinguishable as distinct populations.

Theoretical investigations of the astrophysical behavior of dark matter employ large computer simulations of the evolution of structure in the universe and the galaxies within it. Those simulations take two forms: simulations with only dark matter moving under the influence of gravity and hydrodynamical simulations, which add baryonic physics such as radiation, star formation, and supernova explosions. Because of lower computational demands, simulations with only dark matter can reach higher resolution than hydrodynamical simulations, which have just recently begun to produce galaxies that closely resemble real galaxies. In both cases, the results can be compared with observations to evaluate whether the physics assumed in the simulations is correct.

Gravitationally bound concentrations of dark matter are referred to as halos, although as with much astronomical terminology, the name is misleading because dark-matter halos have high central mass densities and lower densities in their outskirts. If dark matter is cold, one prominent prediction from simulations is that large dark-matter halos should be surrounded by enormous numbers of smaller halos. The population of dark-matter halos follows a characteristic distribution in mass,[7] known as the mass function, in which the number of satellite dark-matter halos $N$ scales with the mass of the host galaxy $M$ as $dN/dM \sim M^{-1.9}$.

According to the mass function, a massive galaxy like the Milky Way should be accompanied by many smaller dark-matter halos. But hydrodynamical simulations suggest that only the halos with masses above about $10^8$ solar masses ($M_\odot$)—four orders of magnitude smaller than the Milky Way's dark-matter

halo—can form stars. Although halos below that limit may be detectable via gravitational lensing or their dynamical effects on thin streams of stars orbiting the Milky Way, most recent efforts to constrain dark-matter properties focus on the smallest concentrations with visible counterparts: dwarf galaxies.

Using the mass function to compare the observed dwarf-galaxy population around the Milky Way with theoretical expectations provides some of the strongest constraints on dark matter. The first such comparisons in the late 1990s and early 2000s revealed a major discrepancy: The number of dwarf galaxies around the Milky Way (11 as of 2005) was more than an order of magnitude short of the $\Lambda$CDM expectation.[8] In an excellent example of scientific branding, the mismatch was labeled the "missing satellite problem," and it provided significant motivation for considering alternative models of dark matter, including warm dark matter and fuzzy dark matter (see box 2). It now appears that the problem lay with the observational searches for dwarf galaxies.

Since 2005 the rapid progress of digital surveys in covering the sky has increased the total number of observed dwarfs orbiting the Milky Way to nearly 60, but all regions of the sky have not been searched with the same sensitivity. When that incompleteness is accounted for, the observed number of dwarf galaxies places strict limits on dark-matter-particle properties. Dark-matter models in which the predicted number of dark-matter halos is smaller than the observed number of dwarfs can be ruled out. Warm dark matter and self-interacting dark-matter models produce fewer dark-matter halos at $10^8 \, M_\odot$, and therefore fewer dwarf galaxies, than CDM. With the latest dwarf-galaxy mass measurements, state-of-the-art theoretical models exclude warm dark-matter particles with masses below 6.5 KeV. The measurements also place strong constraints on other theories of dark matter.[9]

Future surveys, including with the Vera C. Rubin Observatory, scheduled for first light in 2023, are expected to approximately double the current Milky Way dwarf-galaxy satellite population over the coming decade. As of now, the population appears to be consistent with $\Lambda$CDM predictions[10] down to about $10^8 \, M_\odot$. Mass measurements for the dozens of dwarfs Rubin will likely discover will make the test more sensitive, thereby strengthening limits on non-CDM dark-matter scenarios, such as fuzzy dark matter and dark matter that can interact with standard model particles.[11]

## Light from dark matter

CDM particles are not expected to interact with baryons, but they may occasionally interact with each other. In currently favored models, dark-matter particles have masses above 1 GeV and can collide and annihilate into familiar standard-model species. The large particle mass means that the annihilations would produce high-energy photons, typically at gamma-ray wavelengths. Searches for that gamma radiation are called indirect-detection experiments because they seek particles originating from dark matter rather than dark matter itself.

The rate of dark-matter annihilation is proportional to the dark matter's density squared. Thus the brightest sources of annihilation radiation will be the nearest and densest concentrations of dark matter. The prime indirect-detection target is the Milky Way's center,[12] which is just 25 000 light-years away. Frustratingly, though, the galactic center also hosts every other known astrophysical source of gamma rays, including supernova remnants, pulsars, and occasional accretion onto the Milky Way's central black hole.

The next-best locations to look for dark-matter annihilation are the Milky Way's dwarf galaxies. They are relatively nearby and are free of any known sources of gamma rays, making them remarkably clean targets for detecting high-energy radiation from annihilating dark-matter particles. As with other dark-matter tests, stellar spectroscopy is critical—velocity-dispersion measure-

### BOX 1. STAR CLUSTER OR GALAXY?

Astronomers classify gravitationally bound systems containing stars into two categories: galaxies and star clusters. Both can include either young or old stars, but galaxies encompass a broader range of masses, sizes, and morphologies. The fundamental difference between the two is thought to relate to their dark-matter content. Galaxies, such as Leo IV (left image), form in deep gravitational potential wells established by concentrations of dark matter and contain at least five times as much dark matter as ordinary matter. Clusters, such as Palomar 12 (right image), arise from unusually dense gas clouds and do not contain detectable amounts of dark matter. Despite having a similar luminosity to Palomar 12, Leo IV is invisible in the image below because the surface density of its stars is smaller by a factor of 100.

Before the discovery of ultrafaint dwarf galaxies, the two classes could be separated using the classic Potter Stewart aphorism "I know it when I see it." Yet as increasingly faint dwarf-galaxy candidates were identified, the properties of star clusters and the new galaxies overlapped. A more rigorous definition was needed. In 2012 Beth Willman and Jay Strader proposed the following: "A galaxy is a gravitationally bound collection of stars whose properties cannot be explained by a combination of baryons and Newton's laws of gravity."[18] Their definition avoids specifically referring to dark matter, as its existence has not yet been confirmed, and enables a stellar system to be classified by comparing its dynamical mass with the mass of its stars.

Ultrafaint dwarf galaxy 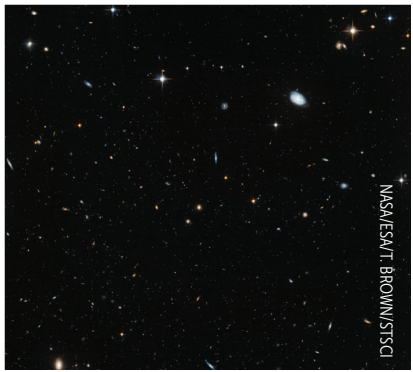   Star cluster 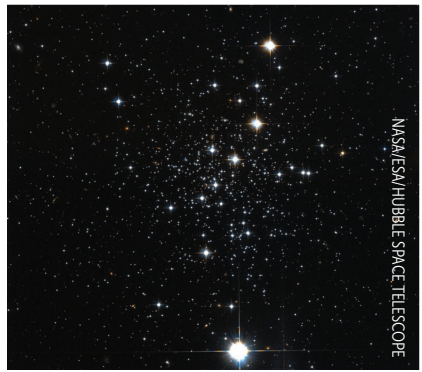

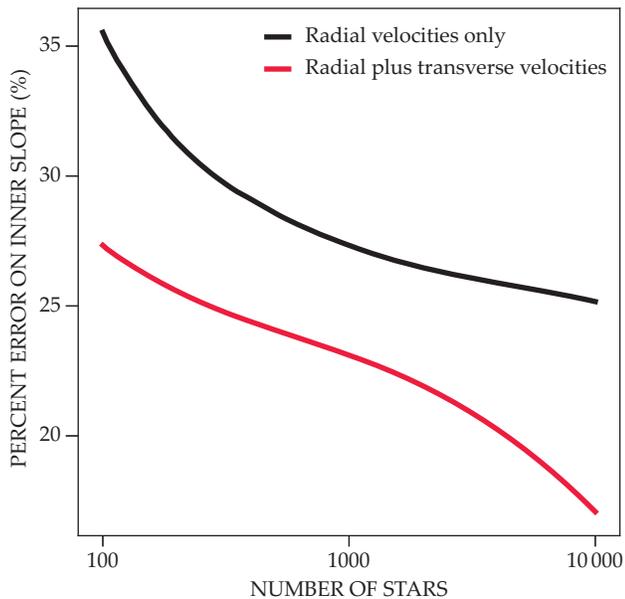

**FIGURE 3. THE PRECISION** to which a galaxy's mass density profile can be determined depends on the number of observed stars. Here, the estimated error in the key property of the density profile—its inner power-law slope—is plotted as a function of the number of stars observed. The black curve assumes that only radial-velocity measurements are available for each star. The red curve assumes that both radial- and transverse-velocity measurements are available. Errors of 2 km/s are assumed on individual radial-velocity measurements, whereas 5 km/s is assumed for transverse velocities. An inner-slope measurement with a certainty of 4–5 standard deviations can only be obtained with three-dimensional velocity measurements. (Courtesy of Juan Guerra, Yale University.)

ments determine the central dark-matter density, and hence the expected annihilation rate, in each dwarf. Multiple gamma-ray telescopes are searching for annihilation radiation from the Milky Way's satellite galaxies. Although they have yet to convincingly detect dwarf-galaxy gamma-ray emission, their nondetections place the most stringent limits to date on the dark-matter particle's interaction cross section. Specific limits depend on the dark-matter model and annihilation channel being considered, but the experimental sensitivity has reached the theoretically expected cross sections[13] for dark-matter masses lighter than about 100 GeV.

Gamma-ray observations of the Milky Way's dwarf galaxies, especially the numerous and nearest ultrafaint dwarfs, may hold the key to understanding the nature of dark matter. Further improvements to dwarf-galaxy density estimates and more sensitive gamma-ray observations will provide progressively stricter constraints on, and perhaps even a detection of, the elusive matter.

## Cusps versus cores

The third constraint on the nature of dark matter provided by the Milky Way's dwarf galaxies comes from measuring their internal mass distributions. Simulations consistently predict[14] that dark-matter halos consisting of CDM particles have central density distributions where the density $\rho$ as a function of radius $r$ approximately follows $\rho(r) \propto r^{-1}$. Such a dark-matter profile is referred to as "cuspy." Dark-matter particles with larger self-interaction cross sections or smaller masses—as used in warm-dark-matter models—generally lead to less-dense central regions. In those cases, the inner density profile either increases more gradually toward the center or is independent of radius. The flatter density profiles are called "cores."

Observations of dark-matter density profiles in bright dwarf galaxies—with masses of $10^{10}$–$10^{11}$ $M_\odot$—beyond the Milky Way suggest that dark matter is not as centrally concentrated as predicted by ΛCDM. That is, the central dark-matter densities are often core-like and do not increase as rapidly as $r^{-1}$. That discrepancy is known as the "cusp–core problem,"[15] and it could be a signal that dark matter is not cold or that dark-matter particles have a significant interaction cross section.

It has become clear over the past decade, however, that massive dwarf galaxies are not the pristine dark-matter laboratories that astronomers once thought them to be. Their baryon fractions can be as high as about 50% in their central regions. Hydrodynamical simulations of galaxy formation have revealed that when many supernova explosions occur close together in time, enormous quantities of gas are first blown out of the galaxy and later recondense, producing strong fluctuations in the gravitational potential.[16] Even if the dark matter does not directly couple to the baryons, it must respond gravitationally by spreading out and reducing the central dark-matter density.

Those simulation results have intensified interest in galaxies containing fewer stars because those galaxies never host enough supernovae to experience such dark-matter rearrangement. By measuring the central dark-matter distribution in those truly dark-matter-dominated systems, researchers hope to determine whether the density profiles of pure dark-matter halos support the ΛCDM paradigm.

Unfortunately, radial velocities from stellar spectroscopy alone have proven to be insufficient to determine dark-matter density profiles of the Milky Way's dwarf galaxies. Analyses by separate groups over the past decade have favored cored profiles, central cusps, and everything in between. Even when relying on the same stellar data sets, independent teams have been unable to converge on a consistent answer.

The fundamental challenge is that stars orbit in three dimensions around the galaxy in which they reside, but radial velocities constrain only one of the three. Astronomers often make assumptions about the missing two dimensions, such as presuming that the stellar motions are isotropic. But the lack of three-dimensional velocity information leads to a degeneracy between the mass distribution and the stellar orbits. To directly determine each star's orbit—which would be a real breakthrough—radial velocities must be combined with velocities along the other two dimensions of the stars' motions.

## Future observations in 3D

Although stellar motions are inherently 3D, for astronomical purposes it is convenient to divide them into two categories: radial motion along the line of sight and proper motion in the plane of the sky. Radial velocities are obtained from stellar spectroscopy, as described above, whereas proper motions are measured by determining the angular position of a star as a function of time. The ability to test dark-matter models by studying dwarf galaxies is limited by the number of stars that can be observed in each galaxy and the accuracy with which the individual stars' motions can be measured (see figure 3).

> **BOX 2. DARK-MATTER MODELS**
>
> Astrophysicists concluded decades ago that cold dark matter (CDM) best describes the universe. That conclusion was based on comparisons between the observed large-scale distribution of galaxies and numerical simulations of nonlinear gravitational clustering. Yet a particle with the expected CDM properties has failed to materialize in either collider experiments or sensitive underground searches, so theorists have more recently begun to seriously consider other ideas for the nature of dark matter.
>
> In warm-dark-matter models, the particle—perhaps a fourth type of neutrino—has a much smaller mass and moves at significantly higher velocities than putative CDM particles. Such a particle would preclude structure on small scales, so a universe dominated by warm dark matter would contain fewer of the smallest dwarf galaxies.
>
> Another prominent alternative dark-matter scenario is self-interacting dark matter, in which the particles interact often with one another. That interaction could be an analogue of electromagnetic interactions among ordinary matter. As with warm dark matter, the interactions would reduce the number of dwarf galaxies and potentially decrease the dark-matter density at the centers of galaxies.
>
> Fuzzy, or ultralight, dark matter has recently garnered increased interest. In that model, the particles have such small masses ($10^{-22}$–$10^{-19}$ eV) that their de Broglie wavelengths are approximately 1 kiloparsec, which is comparable to the sizes of galaxies and would lead to quantum interference effects on galactic scales.
>
> The detailed implications of the alternative models are still being explored, but prospects for conclusive astrophysical tests in the foreseeable future appear promising.

Currently only the radial velocities of stars in the Milky Way's satellite galaxies can be measured with enough accuracy to test dark-matter models. In bright dwarf galaxies, like those discovered in older photographic surveys, several thousand stars are easily observable with existing telescopes. That is sufficient to determine the stellar velocity dispersion in the galaxy. On the other hand, the most recently discovered ultrafaint galaxies have as few as 5–10 stars bright enough for radial-velocity measurements.

Several new survey instruments that are either in the planning process or beginning operations will include spectrographs that can obtain data for thousands of stars at a time. Those tools, such as the Prime Focus Spectrograph on the 8.2 m Subaru Telescope in Hawaii and the Dark Energy Spectroscopic Instrument on the 4 m Mayall Telescope in Arizona, may produce larger gains for bright dwarfs than for faint ones. Larger future telescopes may bring more of the stars in ultrafaint dwarfs within reach and allow astronomers to measure radial velocities for perhaps hundreds of stars in those systems. Those facilities will be crucial for studying the ultrafaint dwarfs expected to be discovered in the coming years by the Rubin Observatory.

Much as the past 30 years have seen significant improvements in the ability to measure stellar radial velocities, the next 10 years will likely see similar advances to proper-motion measurements. Distant stars' proper motions are determined by comparing their positions with those of stationary background sources in images taken at least a few years apart. Even over decades, stars' transverse motions are extraordinarily tiny—akin to watching human hair grow from the Moon. Still, the *Gaia* satellite, launched by the European Space Agency in 2013, has now measured transverse motions for a remarkable 1.8 billion stars in the Milky Way and small samples of stars in its satellite dwarf galaxies.[17]

Although transverse-velocity measurements represent a phenomenal technical achievement, those acquired so far in the Milky Way's dwarf galaxies remain less than one-tenth as precise as currently available radial velocities and thus do not yet sufficiently constrain the dark-matter distribution. Anticipated ground- and space-based improvements are likely to change that situation in the foreseeable future. The European Space Agency recently released a highly anticipated third data set from *Gaia* that improves the proper-motion precision for individual dwarf-galaxy stars by a factor of two to three, and substantial gains are expected in the next few years. Further in the future, existing *Hubble Space Telescope* images of dwarfs will be combined with observations by the next generation of large telescopes. The exquisite angular resolution of those combined data sets will enable proper-motion measurements for hundreds of dwarf-galaxy stars with velocity errors comparable to current radial-velocity errors.

Determining the dark-matter distribution in dwarf galaxies with sufficient precision to meaningfully test theories of dark matter will ultimately require a combined analysis that features both radial and transverse stellar motions. The full 3D orbits for the observed stars in a dwarf galaxy will reveal the galaxy's underlying gravitational potential, thereby confirming or refuting the $\Lambda$CDM prediction of a cuspy dark-matter density profile and improving constraints on dark-matter annihilation rates. In that way, measurements of infinitesimal motions in seemingly insignificant galaxies may hold the key to determining the nature of dark matter throughout the universe. If so, the Milky Way's entourage may earn its own starring role.

PT